\documentclass[prl,twocolumn,superscriptaddress]{revtex4}
\usepackage{amsmath,graphicx,bm,float}

\usepackage{color}
\usepackage{comment}

\newcommand{\ket}[1]{|{#1}\rangle}
\newcommand{\bra}[1]{\langle{#1}|}

\newcommand{\rA}{\mathrm{A}}
\newcommand{\rB}{\mathrm{B}}
\newcommand{\rP}{\mathrm{M}}

\newcommand{\rS}{\mathrm{S}}

\newcommand{\CNOT}{{\sc cnot} }
\newcommand{\cnot}{{\sc cnot}}

\begin{document}
\title{ 
 Complete and Deterministic Bell State Measurement Using Nonlocal Spin Products}
\author{Keiichi Edamatsu}
\email{eda@riec.tohoku.ac.jp}
\affiliation{Research Institute of Electrical Communication, Tohoku University,  Sendai 980-8577, Japan}

\date{\today}

\begin{abstract}
A simple protocol for complete and deterministic Bell state measurement is proposed.
It consists of measurements of nonlocal spin product operators 
with the help of shared entanglement as an ancillary resource.
The protocol realizes not only nonlocal Bell state measurement between 
a pair of distant qubits
but also a complete Bell filter
that transmits either one of the Bell states indicated by the measurement outcome.
These schemes will be useful in quantum technologies
where nonlocal Bell state measurement is indispensable.
\end{abstract}


\maketitle

\section{Introduction}
\begin{figure}[b]
\includegraphics[width=55mm]{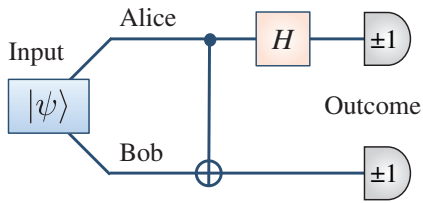}
\caption{
A simple circuit model for the Bell measurement composed of a \CNOT gate followed by a Hadamard gate \cite{Nielsen00a}.
\label{fig.classicbellmeas}
}
\end{figure}

Bell state measurement, or Bell measurement, is an essential concept in quantum technologies  \cite{Nielsen00a}.
The most simple Bell measurement scheme is illustrated in Fig.~\ref{fig.classicbellmeas}.
Although this scheme is simple 
and instructive, 
it requires a nonlocal controlled-not (\cnot) operation between the two qubits. 
A deterministic \CNOT operation 
and thus 
a complete Bell measurement 
would be possible in various qubit systems where two qubits 
are situated 
close to each other.
However, Bell measurement between distant qubits is not possible 
using only
local operation and classical communication (LOCC).
More realistic implementations of Bell measurement using linear optics \cite{Weinfurter:1994tc}
are sometimes applied to various proof-of-principle demonstrations of  
quantum information protocols 
such as quantum teleportation and entanglement swapping etc.\,\cite{Bouwmeester97a,Pan98a,Goebel:2008cl,Kaneda:2012dg},
but it is known that they cannot be complete and deterministic,
i.e., the linear-optical implementations can only partly distinguish among the four possible Bell states \cite{Weinfurter:1994tc,Kok:2007ep}.

In this letter, 
a simple scheme for complete and deterministic Bell measurement is proposed.
It consists of the measurements of {\it nonlocal spin product operators},
the members of nonlocal product observables \cite{Brodutch:2016jv},
with the help of 
shared entanglement as a resource.
The protocol enables nonlocal Bell measurement between 
a pair of distant qubits.  
Furthermore,  not only 
Bell measurement
but also a complete Bell filter
is realizable, 
where the output state turns out to be either one of the Bell states indicated by the measurement outcome.

\section{Nonlocal Spin Products and Bell Bases}
We consider a bipartite qubit system where
a pair of qubits 
is 
distributed between Alice and Bob.
In general, 
the state vector $\ket{\Psi}$ of the bipartite system is expressed as
\begin{align}
 \ket{\psi} = \sum_{\mu, \nu} c_{\mu\nu}\ket{\mu\nu} , 
\label{eq.system}
\end{align}
where 
$
\ket{\mu\nu} \equiv \ket{\mu}_\rA \otimes \ket{\nu}_\rB
$
and 
$\ket{\mu}_\rA$ ($\ket{\nu}_\rB)$ is the eigenstate of 
the Pauli operator $\sigma_z$ 
on Alice's (Bob's)  site
having eigenvalues $\nu$, $\nu=\pm1$.
Note that
$
\sum_{\mu, \nu} \left| c_{\mu\nu} \right| ^2 =1.
$
Hereafter we sometimes write just $+$ and $-$ for the eigenvalues $+1$ and $-1$, respectively.
For instance,
$
\ket{+-} 
= \ket{+}_\rA \otimes \ket{-}_\rB 
= \ket{+1}_\rA \otimes \ket{-1}_\rB .
$
Also note that
we regard the states $\ket{+}$ and $\ket{-}$ 
as $\ket{0}$ and $\ket{1}$ 
in the standard qubit representation, 
respectively.
The Bell bases are defined as
\begin{align}
\ket{\Phi^\pm} &= \frac{1}{\sqrt2} \left( \ket{++} \pm \ket{--}\right)  ,
\label{eq.BB1}\\
\ket{\Psi^\pm} &= \frac{1}{\sqrt2} \left( \ket{+-} \pm \ket{-+}\right)  .
\label{eq.BB2}
\end{align}
Using the Bell bases, $\ket{\psi}$ in (\ref{eq.system}) is rewritten as
\begin{align}
\ket{\psi} = c_1 \ket{\Phi^+} +c_2 \ket{\Phi^-} + c_3 \ket{\Psi^+} +c_4 \ket{\Psi^-},
\label{eq.system2}
\end{align}
where 
$
c_1=(c_{++}+c_{--})/\sqrt2, 
$
etc.

The simple circuit for the Bell measurement given in Fig.~\ref{fig.classicbellmeas}
transforms the input states
$\ket{\Phi^+}$, $\ket{\Phi^-}$, $\ket{\Psi^+}$ and $\ket{\Psi^-}$
into the output states
$\ket{++}$, $\ket{-+}$, $\ket{+-}$ and $\ket{--}$,
respectively.
Thus 
we can distinguish all the Bell states 
by observing the local measurement outcomes 
on Alice's and Bob's qubits.
However, as noted earlier,
the simple scheme requires nonlocal \CNOT operation between distant qubits.

Nonlocal spin product operators, 
or nonlocal spin products, 
$S_{ij}$ are 
expressed as
\begin{align}
 S_{ij} \equiv \sigma_i \otimes \sigma_j ,
\quad
(i,j = x, y, z)
\end{align}
where the first and second Pauli operators act on Alice's and Bob's qubits, respectively.
The eigenvalues of $S_{ij}$ are $\pm1$ and each 
of them
is doubly degenerate. 
For instance, eigenvectors of $S_{zz}$ for 
eigenvalues $m=+1$ and $-1$ are written as
\begin{align}
\ket{m=+1} &= 
  \alpha \ket{\Phi^+} + \beta \ket{\Phi^-} , 
\label{eq.evzz1}  \\
\ket{m=-1} &= 
  \gamma \ket{\Psi^+} + \delta \ket{\Psi^-}.   
\label{eq.evzz2} 
\end{align}
Importantly, we find
\begin{align}
 [S_{ii}, S_{jj}] =0 .
\end{align}
Thus, for instance, $S_{zz}$ 
commutes 
with $S_{xx}$.
As a result,  $S_{zz}$ and $S_{xx}$ are compatible having a complete 
orthonormal set of common eigenbases:
\begin{align}
\ket{m=+1, n=\pm1} &= 
\ket{\Phi^\pm},
\label{eq.SPB1} \\
\ket{m=-1, n=\pm1} &= 
\ket{\Psi^\pm},
\label{eq.SPB2}
\end{align}
where $n$ refers to the eigenvalue of $S_{xx}$.
These are nothing 
other 
than the Bell bases (\ref{eq.BB1}) and (\ref{eq.BB2}).
Thus, by observing $S_{zz}$ and $S_{xx}$ for a given input state, we carry out a complete Bell measurement.

\section{Measurement of the Spin Products} 
Suppose Alice and Bob want to measure any 
nonlocal spin product $S_{ij}$
for an arbitrary system state $\ket{\psi}_\rS$ expressed in (\ref{eq.system}) or (\ref{eq.system2}).
Here, the suffix S after the state vector 
refers to the system to be measured.

Measuring a component of $S_{ij}$ is simple.
Since the measurements of $\sigma_i$ 
by
Alice and $\sigma_j$ 
by 
Bob are compatible, 
we can make simultaneous local measurements of $\sigma_i$ and $\sigma_j$, 
and then 
compute the product 
of Alice's and Bob's outcomes to obtain the measurement result of $S_{ij}$.
Consider, for example, the measurement of $S_{zz}$.
The measurement operators $M{(\mu\nu)}$ for
the four possible combinations of 
Alice's and Bob's outcomes, $(\mu,\nu)$, 
are the projective operators:
\begin{align}
M{(\mu\nu)} =\Pi(\mu\nu)=\ket{\mu\nu}\bra{\mu\nu},
\label{eq.localmeasop}
\end{align}
where $\Pi(...)$ is the projector to the state $\ket{...}$.
The corresponding POVM (positive operator valued measure) is 
$E(\mu\nu)=M^\dag(\mu\nu)M(\mu\nu)=\Pi(\mu\nu)$.
Taking a product of $\mu$ and $\nu$, we obtain the outcome $m$ for the measurement of $S_{zz}$.
The POVMs $E_\pm$ for $m=\pm1$ are obtained as
\begin{align}
E_{+} 
= \Pi(++)+\Pi(--) 
= \Pi(\Phi^+)+\Pi(\Phi^-) ,
\label{eq.povm1} \\
E_{-} 
= \Pi(+-)+\Pi(-+) 
= \Pi(\Psi^+)+\Pi(\Psi^-) .
\label{eq.povm2}
\end{align}
We see that $E_\pm$ correspond to the projection to the eigenspaces of $m=\pm1$ presented in (\ref{eq.evzz1}) and (\ref{eq.evzz2}), respectively.

However, this local measurement strategy is inappropriate to make simultaneous measurements of
two or more components of $S_{ij}$, 
for instance, $S_{zz}$ and $S_{xx}$.
Since the above-mentioned measurement of $S_{zz}$ projects the system state to either of the local product staltes
$\ket{++}$, $\ket{--}$, $\ket{+-}$ or $\ket{-+}$,
the succeeding measurement of $S_{xx}$ is no longer identical to that of the original input state.
In other words,
the local projective measurement of $\sigma_z$ on either Alice's or Bob's qubit is not compatible with $S_{xx}$.

\begin{figure}[t]
\includegraphics[height=36mm]{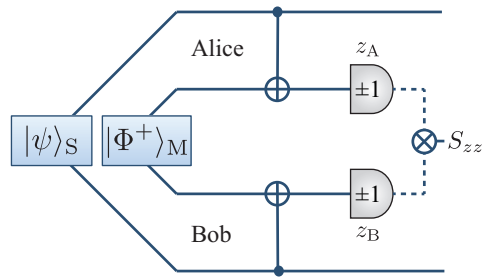}
\caption{
A scheme for the measurement of nonlocal spin product $S_{zz}$.
\label{fig.nlmeas}
}
\end{figure}

Another strategy for the measurement of spin products is the {\it nonlocal} measurement 
making use of additional entanglement shared by Alice and Bob.
Suppose Alice and Bob share a maximally entangled Bell state (ebit):
\begin{align}
\ket{\xi}_\rP &= 
\ket{\Phi^+}_\rP .
\label{eq.meter}
\end{align}
The suffix M indicates that it is used as a meter to measure the system state.
Alice and Bob use each qubit in $\ket{\xi}_\rP$ as a meter (probe) to measure their system qubit. 
To do so, each of them makes a \CNOT gate between her/his qubits, as shown in Fig.~\ref{fig.nlmeas},
 and then makes a projective $\sigma_z$ measurement on her/his meter qubit.
Note that when her/his initial meter qubit was fixed as $\ket{+}$, 
the measurement would be the projective local measurement of $\sigma_z$.
After the \CNOT gates, the initial state 
$
\ket{\psi}_\rS \otimes \ket{\Phi^+}_\rP
$
is converted to
\begin{align}
\ket{\psi}_\rS \otimes \ket{\Phi^+}_\rP &\to 
 \left(c_1 \ket{\Phi^+} + c_2 \ket{\Phi^-}\right)_\rS \otimes \ket{\Phi^+}_\rP \nonumber \\
&\quad + \left(c_3 \ket{\Psi^+} + c_4 \ket{\Psi^-}\right)_\rS \otimes \ket{\Psi^+}_\rP.
\label{eq.nlmeas1}
\end{align}
Let Alice's (Bob's) outcome be $z_\mathrm{A}$ ($z_\mathrm{B}$).
The first term of the right hand side of (\ref{eq.nlmeas1}) 
corresponds to the case where the measurement outcome is $(z_\mathrm{A}\, z_\mathrm{B})=(++)$ or $(--)$,
while the second term to $(+-)$ or $(-+)$.
By simply taking a product of the local meter outcomes of Alice and Bob,
$z_\mathrm{A}z_\mathrm{B}=m=\pm1$ is obtained 
and thus the measurement of $S_{zz}$ is complete.
The measurement operators $M_\pm$ for $m=\pm1$ are
\begin{align}
M_+ &= 
 \Pi(\Phi^+)+\Pi(\Phi^-) ,
\\
M_- &= 
 \Pi(\Psi^+)+\Pi(\Psi^-) .
\label{eq.nlmeasop}
\end{align}
The corresponding POVMs are identical to that presented in (\ref{eq.povm1}) and (\ref{eq.povm2}).
$M_\pm$ project the system state to the eigenspaces of $m=\pm1$ presented in (\ref{eq.evzz1}) and (\ref{eq.evzz2}), respectively.
Note that
the system state is still a superposition of 
$\ket{\Phi^+}$ and $\ket{\Phi^-}$
(or $\ket{\Psi^+}$ and $\ket{\Psi^-}$),
preserving sufficient information 
for the succeeding 
$S_{xx}$ measurement.

\begin{figure}[t]
\hspace{5mm} (a)\hspace*{\fill} \\
\includegraphics[height=36mm]{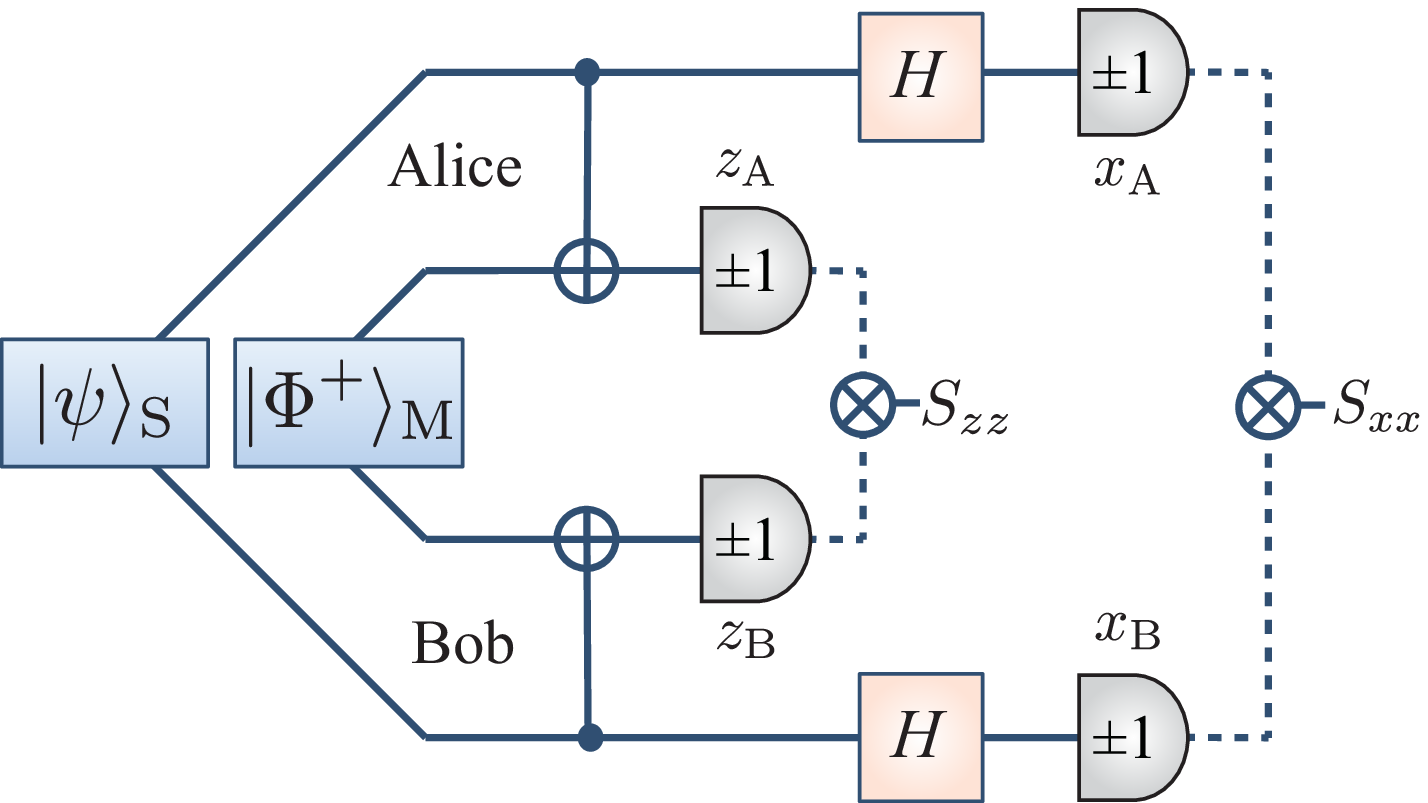}
\vspace{3mm}\\
\hspace{5mm} (b)\hspace*{\fill} \\
\includegraphics[height=36mm]{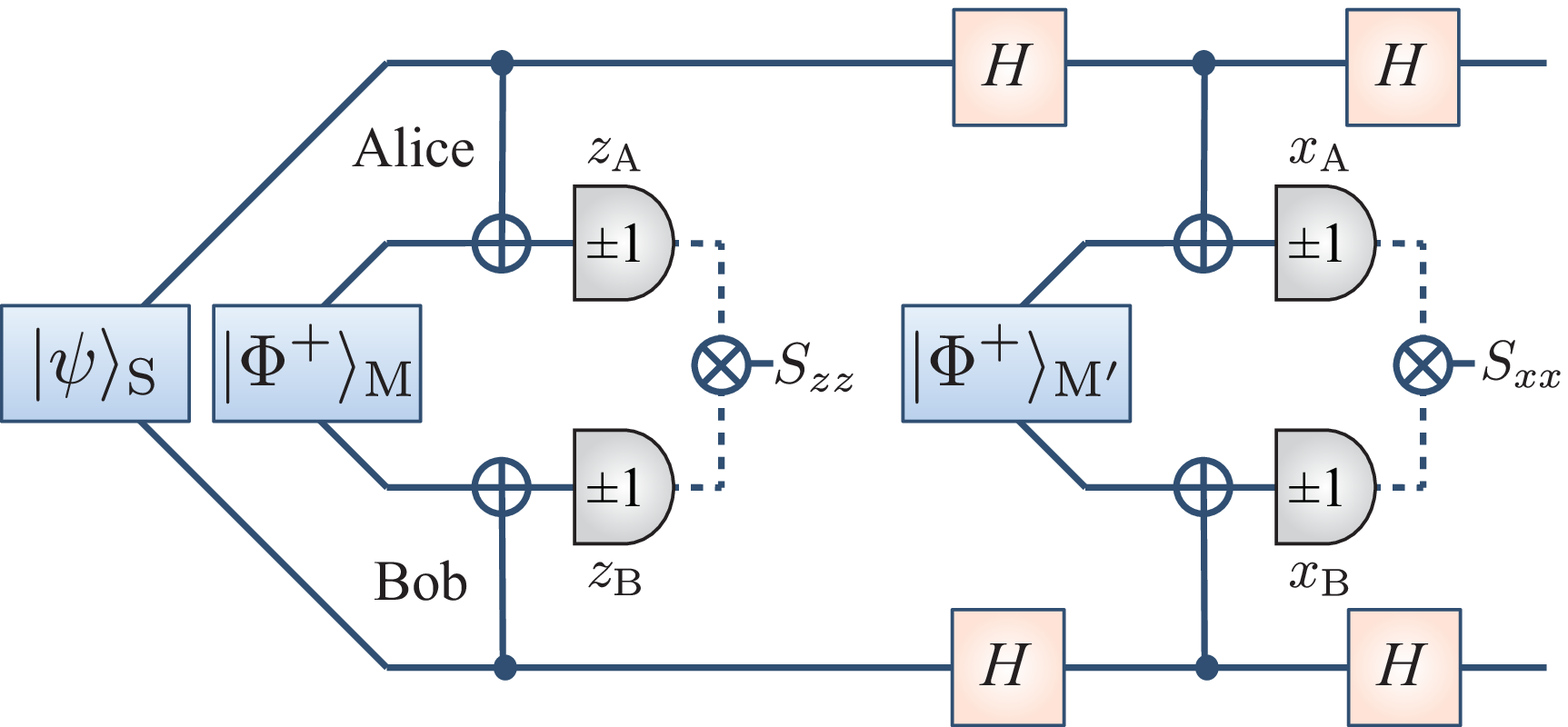}
\caption{
Two schemes for the complete Bell measurement making use of sequential spin product measurements,
$S_{zz}$ and $S_{xx}$.
(a) With nonlocal $S_{zz}$ and local $S_{xx}$ measurements.
(b) With nonlocal measurements of $S_{zz}$ and $S_{xx}$.
Note that the scheme (b) acts as a complete Bell filter.
\label{fig.bellmeas1}
}
\end{figure}

\section{Complete Bell Measurement and a Bell Filter} 
Subsequent to
the nonlocal measurement of $S_{zz}$ described above,
Alice and Bob make a $S_{xx}$ measurement 
on
the system state.
As shown in 
(\ref{eq.SPB1}) and (\ref{eq.SPB2}), 
the measurement of $S_{xx}$ combined with 
$S_{zz}$
discriminates 
$\ket{\Phi^+}$, $\ket{\Phi^-}$,
$\ket{\Phi^+}$, or $\ket{\Phi^-}$ from (\ref{eq.nlmeas1}).
In this way, 
a complete Bell measurement can be carried out.

For the $S_{xx}$ measurement,
either local or nonlocal strategy can be used.
If only the measurement outcome 
matters,
the simple local strategy shown in Fig.~\ref{fig.bellmeas1}\,(a) is appropriate.
In this case, 
the POVMs
$E_{mn}$, 
where suffixes $m$ and $n$ indicate the measurement outcomes of  
the preceding $S_{zz}$ and the following $S_{xx}$, respectively,  
are written as
\begin{gather}
E_{++} =  \Pi(\Phi^+),  \ 
E_{+-} = \Pi(\Phi^-),  
\label{eq.povm3}\\
E_{-+} =  \Pi(\Psi^+),  \ 
E_{--} =  \Pi(\Psi^-). 
\label{eq.povm4}
\end{gather}

On the other hand,
if  the system state at the output should be 
preserved 
in one of the resultant eigenstates given in  (\ref{eq.SPB1}) and (\ref{eq.SPB2}), 
i.e., one of the Bell bases,
Alice and Bob can use nonlocal strategy at a cost of an additional ebit,  as shown in Fig.~\ref{fig.bellmeas1}\,(b).
In this case, the measurement operators  $M_{mn}$ 
are found to be
\begin{gather}
M_{++} =  \Pi(\Phi^+),  \ 
M_{+-} = \Pi(\Phi^-),  
\label{eq.nlmeasop3} \\
M_{-+} =  \Pi(\Psi^+),  \ 
M_{--} =  \Pi(\Psi^-). 
\label{eq.nlmeasop4}
\end{gather}
Again, the corresponding POVMs are identical to that presented in (\ref{eq.povm3}) and (\ref{eq.povm4}).
Thus, the system state is projected into one of the Bell basis depending on the measurement outcomes.
This procedure functions as a complete {\it Bell filter},
where the output state will be either one of the Bell bases 
indicated by the measurement outcome.

It is noteworthy that, in both strategies, 
all of the outcomes 
are deterministically obtained
and thus the Bell measurement proposed here is complete and deterministic,
at the cost of requiring
one (for the Bell measurement)  or two (for the Bell filter) ebit(s) as a resource.

\begin{figure}[t]
\includegraphics[height=50mm]{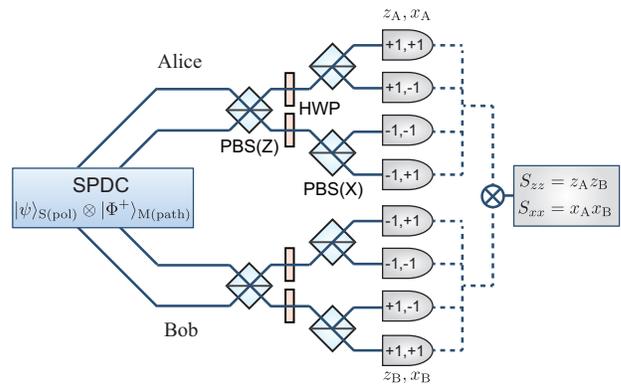}
\caption{
Proposed optical implementation of the complete Bell measurement scheme shown in Fig.~\ref{fig.bellmeas1}\,(a).
Photon pairs are generated by spontaneous parametric down-conversion (SPDC) so that
their path qubits are entangled in $\ket{\Phi^+}_\mathrm{M}$.   
A polarization beamsplitter (PBS) works as a \CNOT gate between the polarization (system) and the path (meter) qubits.
The outcomes of $\sigma_z$
($z_\mathrm{A}$ and $z_\mathrm{B}$ on Alice's  and Bob's qubits, respectively)
are encoded in the output paths of PBS(Z).
After passing through the half-wave plate (HWP), 
which acts as a Hadamard gate,
the outcomes of $\sigma_x$
($x_\mathrm{A}$  and $x_\mathrm{B}$)
are encoded in the output paths of PBS(X).   
\label{fig.photonicexp}
}
\end{figure}

\section{Proposed Experiments}
The measurement schemes described above are applicable to
any physical qubits between which we can prepare entanglement and a \CNOT operation.
However, 
in cases where we can directly make a nonlocal \CNOT operation between qubits held in Alice and Bob,
we could employ a simpler scheme as shown, for instance, in Fig.~\ref{fig.classicbellmeas}
to implement the Bell measurement.
Nevertheless, our scheme is still useful 
when we are not able to use nonlocal \cnot, 
or when we need the function of the Bell filter as well as the Bell measurement. 

Another situation where our schemes may be useful is 
the case of photonic qubits.
It is known that 
with linear optics we cannot implement deterministic \CNOT gates between individual photonic qubits \cite{Kok:2007ep}.
As a result, to date, we could not implement deterministic Bell measurement with linear optics.
Nevertheless, employing the scheme described in this paper we will be able to implement 
the deterministic and complete Bell measurement between photonic qubits.

Suppose we provide a pair of photons to Alice and Bob, 
ss shown in Fig.~\ref{fig.photonicexp}.
The photons' polarizations 
constitute 
the system state $\ket{\psi}_\rS$ of 
interest. 
In order to measure the nonlocal spin products on their polarizations, 
we prepare their path degrees of freedom, i.e., path qubits, in the maximally entangled Bell state $\ket{\Phi^+}_\rP$. 
Entanglement in the path degrees of freedom 
could be directly generated by 
spatial entanglement between photons generated by parametric down-conversion \cite{Shimizu:2008et},
or could be converted from time-bin entanglement \cite{Brendel99a}.
When the polarization qubits are also entangled, it is called 
a hyperentangled state \cite{Barreiro05a}. 
Between the polarization qubit and the path qubit, Alice and Bob employ \CNOT gates using polarizing beamsplitters (PBS).
Thus, the nonlocal measurement of $S_{zz}$ on the photons' polarization qubits is implemented
and
the measurement outcomes are encoded in photons' output paths. 
Then Alice and Bob carry out the local $\sigma_x$ measurement for their polarization qubits using,
for instance, 
two additional PBSs.
This part implements the local measurement strategy of $S_{xx}$. 
At the last stage, Alice 
detects her photon at one of her four output paths, 
as does Bob  at one of his four output paths. 
From the path information, they know the result of $S_{zz}$ and $S_{xx}$,
and thus the complete Bell measurement is carried out in a deterministic way.
One drawback of this linear optics implementation is that
we use an ebit implemented in the path degree of freedom of the photon pair.
As a result, it is difficult to apply this method to Bell measurement between independent photons 
as in a case of quantum teleportation. 
Nonetheless, this method will be useful in many situations of quantum technologies 
where nonlocal Bell measurement is an essential resource.

\section{Conclusions}
In this letter
we see that 
complete and deterministic Bell measurement 
is possible 
in terms of nonlocal spin product measurements.
Although the Bell measurement requires an ebit as a resource,
the scheme is readily realizable using present technologies.
In particular, we propose an optical implementation using linear optics. 
In addition, a complete Bell filter, which requires a couple of ebits and thus seems a bit more difficult to implement,
is also possible.
These schemes will be useful in quantum technologies
where nonlocal Bell measurement is indispensable.

Furthermore,  
the measurement protocol of nonlocal spin products can be extended to measurements 
at weak and any intermediate measurement strength \cite{Brodutch:2016jv}.
Thus, it would be possible to realize generalized measurements of nonlocal spin products
and Bell measurement at any measurement strength.
In this context, 
strength-variable measurements of photon polarization 
and the measurement uncertainty relations have been demonstrated 
\cite{Lund:2010cn,Rozema:2012bx,Baek:2013fh,Ringbauer:2014gk,Kaneda:2014em,Edamatsu:2016ep}.
By extending the protocols described here,
it would be possible  to 
explore measurement uncertainty relations 
in the nonlocal product observables.

\section*{Acknowledgments}
The author is grateful to M. Sadgrove and P. Vidil for valuable discussions.

\vfill

%


%

\end{document}